# IS THE MULTIVERSE HYPOTHESIS CAPABLE OF EXPLAINING THE FINE TUNING OF NATURE LAWS AND CONSTANTS? THE CASE OF CELLULAR AUTOMATA

Francisco José Soler Gil[1] – Manuel Alfonseca[2]## Abstract

The objective of this paper is analyzing to which extent the multiverse hypothesis provides a real explanation of the peculiarities of the laws and constants in our universe. First we argue in favor of the thesis that all multiverses except Tegmark's «mathematical multiverse» are too small to explain the fine tuning, so that they merely shift the problem up one level. But the «mathematical multiverse» is surely too large. To prove this assessment, we have performed a number of experiments with cellular automata of complex behavior, which can be considered as universes in the mathematical multiverse. The analogy between what happens in some automata (in particular Conway's «Game of Life») and the real world is very strong. But if the results of our experiments can be extrapolated to our universe, we should expect to inhabit −in the context of the multiverse− a world in which at least some of the laws and constants of nature should show a certain time dependence. Actually, the probability of our existence in a world such as ours would be mathematically equal to zero. In consequence, the results presented in this paper can be considered as an inkling that the hypothesis of the multiverse, whatever its type, does not offer an adequate explanation for the peculiarities of the physical laws in our world. A slightly reduced version of this paper has been published in the Journal for General Philosophy of Science, Springer, March 2013, DOI: 10.1007/s10838-013-9215-7.**Keywords:** Astrophysics, Cosmology, Multiverse, Fine tuning, Cellular automata## Introduction

The hypothesis that the universe where we live represents, not the whole physical reality, but a particular domain inside a much larger reality which includes many other universes, started to be considered a cosmological possibility in the last decade of the previous century. Although a similar idea was offered in the fifties as a solution to the quantum measurement problem (Everett's «multiple worlds» interpretation), that problem is completely different from those tackled by the cosmologists who uphold the multiverse hypothesis. Therefore, here we will not consider Everett's approach and its subsequent formulations.

---

[1] Universidad de Sevilla y Technische Universität Dortmund,
  e-mail: soler@uni-bremen.de ; solergil@us.es ; francisco.soler@tu-dortmund.de
[2] Universidad Autónoma de Madrid, Escuela Politécnica Superior, e-mail: manuel.alfonseca@uam.es1

The confluence of three different lines of research explains why a hypothesis as speculative and risky as the multiverse has been taken seriously in the latest years. These lines are the following:

(1) Inflationary cosmology.

(2) Different attempts to build a quantum gravity theory.

(3) Research on the effect that a small modification in the structure of physical laws would have on the development of complex beings and life as we know it.

The multiverse question arises in the context of inflationary cosmology. The cosmic inflation hypothesis was proposed initially by Alan Guth in 1981. Guth was trying to explain two phenomena which the standard cosmological model could not explain: (1) the homogeneity of those regions in the universe which had never been able to interact [the so-called horizon problem] and (2) the fact that the universe seems to be approximately flat, which entails that, in the beginning of the expansion, the density parameter of the universe must have had a value extremely near to the critic density [the so-called flatness problem].

Guth proposed that the universe experienced a process of exponential expansion between $10^{-37}$ and $10^{-35}$ seconds after the Big Bang. According to Guth's initial proposal, the reason for this would be given by the value of Higgs's field $\Phi$ as deduced from a certain version of the grand unifying theory. The potential of this field $V(\Phi)$ would act at that time as a cosmological constant, accelerating enormously the expansion of the universe. At the end of this stretch of time, inflation would be replaced by an expansion similar to that described by the standard model, which would still be valid, except for its application to the first stages of the universe.

The inflationary scenario offers an answer to the two mentioned questions. The horizon problem is solved because what today makes our observable universe comes from a very small region with mutual interactions before the exponential expansive phase. And the flatness problem is solved because the universe, as a consequence of the inflation, has reached such dimensions that it appears to be practically flat, even although it may still possess some curvature.

But it so happens that the initial Guth theory, together with several other proposals made to support inflation, have been proved unfeasible, because they do not fit some of the known cosmological parameters; because they predict an universe less homogeneous than what can be observed; because they display internal inconsistencies; and so forth. However, this hypothesis is still appealing for cosmologists, thus new, ever more refined inflationary scenarios have been developed. Currently, the model which appears to present less problems – developed mainly by Andrei Linde, Alex Vilenkin and co-workers– suggests that the cosmic inflationary process never ends, that the universe expands exponentially forever, while here and there different domains are being formed, such as our observable universe, which are regions of a much larger physical reality: regions where the potential pushing the inflation has reached a minimum value, and where therefore the exponential growth of the cosmos has stopped. As these regions are causally disconnected from one another, it seems that, if we accept the inflationary hypothesis, we should also accept the existence of the multiverse.

Another independent line of research which leads to the idea of the multiverse is the search for a quantum gravity theory. It will suffice to say, briefly, that we have at present mainly two different approaches towards this theory: the superstring hypothesis, and the quantum loop theory, both of which end up in the idea of the multiverse. In the case of superstrings, the problem is that, instead of there being a single physical structure complying with the requirements considered fundamental for this frame, there are about $10^{500}$ (or as some say, $10^{1000}$) possible structures. This is a regrettable situation. And the solution proposed by Susskind and others is that physical reality carries out all those possibilities. So we would live in a multiverse where all the possible universes within the frame of string theory are also real universes.

As to the quantum loop hypothesis, it so happens that the first tentative cosmological models being developed inside this frame −Bojowald models− suggest that our universe suffered a process of collapse previous to the Big Bang described by the standard cosmology. This has reinforced Smolin's conjecture that this universe could have resulted from a gravitational collapse inside a larger physical reality. In other words, one universe may give origin to another, inside one of its black holes. Once again, the multiverse scenario.

Finally, the multiverse hypothesis has been proposed as a solution to the problem of the fine tuning of physical constants and laws. This problem can be stated thus: since the eighties, we have a detailed standard model, both in cosmology and in particle physics. This makes it possible to analyze, theoretically and through computer simulations, questions such as the consequences for the cosmos of slight changes in some of the parameters demanded by these models. The result of these researches has been the discovery that a certain number of parameters, both in the standard cosmological model, as in the standard particle physics model, give the impression of being finely tuned, in the sense that, if they had had values minimally different from those they actually have, life in the cosmos −and in many cases every complex structure− would have been physically impossible.

Some authors interpret this fact as an inkling that our universe is nothing but a single domain in a much larger reality, in such a way that we inhabit just that domain of reality where the appropriate conditions for the existence of life as we know it prevail.

The objective of this paper is analyzing to which extent the multiverse hypothesis provides a real explanation of the peculiarities of the laws and constants in our universe. To reach this goal, the paper is divided in the following sections:

In the first section, we will consider some of the examples of fine tuning discussed in the latest years, so as not to lose sight of what we want to explain with the help of the multiverse. In other words, what is the fine tuning of the universe.

In the second section we will discuss the question of which version of the multiverse should be assumed to eliminate fully the problem of the fine tuning of the universe. We shall show that only the «mathematical multiverse» proposed by Max Tegmark prevents the question of fine tuning to appear again in the multiverse context.

In the third section we will suggest a way to test that hypothesis. We shall select cellular automata as examples of possible universes where we can test some of the «predictions» offered by Tegmark as regards the multiverse hypothesis.

In the fourth section we will detail the experiments we have performed with cellular automata with respect to Tegmark's «predictions». They try to show how the behavior of cellular automata is affected by changes in their rules −the equivalent to the laws of nature in a universe− in a relation to their ability to develop complex structures.

In the fifth section we will state the consequences of our study regarding the question of whether the multiverse hypothesis can explain the fine tuning of our universe. Our provisional answer is on the negative: if we accept the multiverse as the explanation of the fine tuning of universe, we should expect that the laws of this universe would be less simple as they are. In particular, it seems that we could expect that at least some of the laws and constants of nature should show a certain time dependence.

## 1. The fact of the fine tuning of the universe

All along the last century, especially in its last decades, a surprising fact about the universe where we live has been discovered: the fact that its architecture possesses very peculiar properties, in the sense that very slight changes in the combination of physical laws and constants of nature would have the consequence that the cosmos would become a physical system hostile to the development of life. We give the name «fine tuning» of the universe to the fact that nature behaves following one of the (at least apparently) scarce hospitable combinations of laws and constants. In order not to leave this exposition in a too abstract plane, we shall mention a few concrete examples of this tuning. These examples have been taken from Robin Collins paper The evidence of fine tuning, one of the clearest presentations of the matter.

(a) The cosmological constant:

> «The smallness of the cosmological constant is widely regarded as the single greatest problem confronting current physics and cosmology. [...] Apart from some sort of extraordinary precise fine-tuning or new physical principle, today's theories of fundamental physics and cosmology lead one to expect [...] an extraordinary large effective cosmological constant, one so large that it would, if positive, cause space to expand at such an enormous rate that almost every object in the Universe would fly apart, and would, if negative, cause the Universe to collapse almost instantaneously back in on itself. This would clearly make the evolution of intelligent life impossible.
>
> What makes it so difficult to avoid postulating some sort of highly precise fine-tuning of the cosmological constant is that almost every type of field in current physics [...] contributes to the vacuum energy. [...] [When] physicists make estimates of the contribution to the vacuum energy from these fields, they get values of the energy density anywhere from higher $10^{53}$ to $10^{120}$ than its maximum life-permitting value»[3].

(b) The strong and the electromagnetic forces:

---

[3] COLLINS (2003) 180-181.

«A 50 percent decrease in the strength of the strong force, for instance, would undercut the stability of all elements essential for carbon-based life, with a slightly larger decrease eliminating all elements except hydrogen»[4]

«[Around] a fourteen-fold increase in the electromagnetic force would have the same effect on the stability of elements as a 50 percent decrease in the strong force»[5].

(c) Carbon production in stars:

«[A] change of more than 0.5% in the strength of the strong interaction or more than 4% in the strength of the Coulomb [electromagnetic] force would destroy either nearly all C or all O in every star. This implies that irrespective of stellar evolution the contribution of each star to the abundance of C or O in the ISM [interstellar medium] would be negligible. Therefore, for the above cases the creation of carbon-based life in our universe would be strongly disfavored»[6]

(d) The proton/neutron mass difference:

«The neutron is slightly heavier than the proton by about 1.293 MeV. If the mass of neutron were increased by another 1.4 MeV −that is, by one part in 700 of its actual mass of about 938 MeV− then one of the keys steps by which stars burn their hydrogen to helium could not occur [...]:

$$p+p \rightarrow deuteron+positron+electron\ neutrino+0.42\ MeV$$

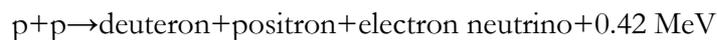

[...] On the other hand, a small decrease on the neutron mass of around 0.5 to 0.7 MeV would result in nearly equal numbers of protons and neutrons in the early stages of the Big Bang [...] resulting in an almost all-helium universe»[7]

Accepting thus that there is a delicate tuning of the laws and physical constants, without which the development of complex chemical structures would not have been possible, especially life (and most especially intelligent life, whose appearance requires doubtlessly that favorable conditions are maintained much longer than what is required by one-cellular life), the question is how to interpret this fact. What does this fine tuning tell us about the physical reality? Is it a meaningful datum, or mere chance? And if the former, what does it entail? What is it pointing at?

In the latest years, some authors suggest that fine tuning is an inkling that the cosmos is much wider than we assumed and is made of domains with different combinations of laws and constants, in which case we must inhabit one of those oasis favorable to life in the middle of a mostly inhospitable physical whole. This is equivalent to propose the multiverse as the explanation of the fine tuning observed in the cosmos.

Now then, under which conditions can the multiverse really explain the fine tuning of our universe? We will tackle this question in the next section.

---

[4] Ibidem 182-183. Taken from BARROW - TIPLER (1986) 326-327. This book from Barrow and Tipler is the very classical exposition on fine-tuning of the universe, and, as such, it is highly recommended for the interested reader.
[5] Ibidem.
[6] Ibidem 185. Cited from OBERHUMMER et al. (2000) 90.
[7] Ibidem 186-187.

## 2. Types of multiverse and their adequacy to explain the fine tuning

In the specialized literature there are several types of entities called «multiverse». The main ones are the following:

1. On the first place, we find authors who give the name of multiverse to the infinite universe, following this reasoning: the observable universe, the environment which includes all the objects whose light has reached us since the Big Bang[8] to our time, has now a radius of about $4 \cdot 10^{26}$ meter; the volume of the corresponding sphere is called «the Hubble volume». It is thus a partial domain inside the infinite universe. Everything we could examine, whatever the power of our telescopes, is included inside our Hubble volume. If there is something beyond, we don't know, and it can never affect us causally. In fact, for us, its existence does not matter. Under a purely empiricist criterion (which we are far from enforcing) we could say that the assertion that there is something beyond the Hubble volume is not even scientific.

   Some authors suggest that we should consider every sphere of the same size as our observable universe as a full-fledged universe. Since an infinite universe would contain infinitely many spheres of this kind, they give it the name of multiverse.

   Actually this terminology is a rather unfortunate choice, because it makes us take as a set of different entities what really makes a single physical system —the open universe— endowed with a high degree of unity, which cannot be decomposed in «sphere-universes» or any other cosmological sub-unities, except in an arbitrary way.

2. The second type of multiverse derives from physical hypotheses which have not reached an empiric support that would allow them to become standard theories, but a certain number of specialists trust that they are adequate to reality, and try to develop them and provide the empiric support they still lack. In this group we may rank the multiverse derived from the eternal inflation scenario proposed by Linde and Villenkin; the multiverse containing the so-called «cosmic landscape», i.e. all the possible realizations of superstring theory; the scenario defended by Smolin of multiple universes generated in black holes; and similar conjectures. In this case, the term multiverse is used to represent something completely new —as against the infinite universe.— In all these scenarios, the various domains differ structurally from one another. This means that the laws of physics can be partially different in each domain, although all of them obey a common general physical structure. On the other hand, each of the cosmic domains is completely —or almost completely— causally disconnected from the others. Therefore each can be considered as an authentic «universe island» which would continue its evolution according to its own dynamics although the remainder of the universe disappeared in an immense cosmic cataclysm.

3. Finally, since a few years ago, the possible existence of a multiverse incomparably larger than the former is being discussed. This idea, proposed by the physicist Max Tegmark, consists in assuming that:

---

[8] Actually we can only receive the light emitted after what is usually called the «surface of final dispersion», the instant when radiation uncoupled from matter. This happened about 100,000 years after the Big Bang. But these details are not important here. After all, 100,000 years are not much... at the cosmological scale.

> «[...] mathematical existence and physical existence are equivalent, so that *all* mathematical structures exist physically as well»[9].

Tegmark's suggestion results in the conception of a multiverse where every possible combination of laws and natural constants occurs in fact in one or another domain of reality. In the cosmic scenario proposed by Tegmark, there are no privileged mathematical structures, nor privileged initial or boundary conditions, nor physical constants of any type whose values are restricted to such and such concrete value. Every consistent mathematical structure is realized. (Or, more precisely, every consistent mathematical structure is a physical universe). The only reason behind the peculiarities of the universe we observe is anthropic: we just observe the world which is consistent with our own existence.

In fact, this third type of multiverse is the only one that provides us with an scenario which does not leave room to the question about the actual values of the physical constants, nor to the question of why the physical laws are what they are. That is, only a multiverse which realizes all the consistent mathematical structures seems a candidate with possibilities to solve the question of the fine tuning of the universe, for in the other multiverses this question appears again, this time in the multiverse frame.

To see that this is so, let us look at a concrete example: the cosmic landscape of string theory may contain about $10^{1000}$ structurally different universes, but they share common features, such as this: all of them possess physical laws of the quantum type. None of these universes may be ruled, for instance, by a Newtonian physics structure. This is an interesting detail, as the huge importance of quantum effects for the appearance of the chemical structures basic for life makes us suspect that, if the multiverse contained only worlds based on variations of classical physics, not one of them would be apt for the existence of life. Therefore, whatever the enormous size of the superstring cosmic landscape, it is still a biophile scenario which suggests design, in contrast with the unrealized possibilities of completely sterile multiverses.

Most of the multiverse models proposed up to now −those of Linde, Vilenkin, Susskind, Smolin, etc.− are subject to this problem: they only realize a too small number of all the possible physical structures. Therefore, the fact that one of them is apt for life is still surprising.

Let us underline this: to notice the limitation of these multiverses, at first sight so vast, what we must do is look at them from the −incomparably larger− perspective of all the mathematical structures which could be considered as the basis for the laws of a possible universe. In other words: in principle, if we start from the set of logically possible universes, it is possible to define in them a great variety of subsets (which would be the possible multiverses), as well as a great variety of mechanisms generating such subsets[10]. And each of these subsets of universes, together with each of their possible generating mechanisms, will possess certain features, more or less favorable for the development of complex structures, living beings, intelligent observers, or any other type of realities. What eventually takes us again to a situation of fine tuning which should be explained. In the words of Stoeger:

---

[9] TEGMARK (2004) 483.
[10] Consult about this, for instance, the thoughts by Ellis, Kirchner and Stoeger about the set of physically possible universes and the different kinds of subsets (multiverses) definable in it. These thoughts can be found in the following papers: ELLIS – KIRCHNER – STOEGER (2003), and STOEGER – ELLIS – KIRCHNER (2004).

«If we do have good evidence, and an adequately specific model for, the multiverse to which our own universe belongs, thus providing some explanation for its bio-friendly characteristics, this would not be a complete −let alone an ultimate− explanation. We would still require an explanation for the existence and bio-friendly character of the multiverse itself (bearing in mind that there is no unique prescription for it) and for the process through which it emerged [...]»[11].

For this reason, many authors have come to the conclusion that the postulate of a kind of multiverse just poses the problem of design in a new plane. In the words of Davies: «Multiverses merely shift the problem up one level»[12]. A conclusion which is based on the following reasoning:

«Each law of laws specifies a different version of the multiverse, and not all multiverses are bound to contain at least one biophilic universe. In fact [...] most multiverses would not contain even one component universe in which all the parameter values were suitable for life. To see this, note that each parameter will have a small range of values −envisage it as a highlighted segment on a line− consistent with biology. Only in universes where all the relevant highlighted segments intersect in a single patch (i.e. all biophilic values are instantiated together) will biology be possible. If the several parameters vary independently between universes, each according to some rule, then for most sets of rules the highlighted segments will not concur»[13].

How can this be solved? In principle, it seems that the only solution is the mathematical multiverse proposed by Tegmark. Evidently, the mathematical multiverse is quite different from the others. In this multiverse there are no mathematical structures (or no particular values of constants) privileged, thus there remains no room for design and choice (or chance). In other words, starting from the physical existence of all mathematical consistent structures, the objections placed (among others) by Davies and Stoeger do not apply.

Let us thus look at the next question: Is the mathematical multiverse a viable explanation of the fine tuning of the laws and natural constants of our universe? Can we do something to test this scenario? Or is this just an unwarranted speculation? We will tackle this in the next section.

## 3. Predictions from the mathematical multiverse hypothesis. Cellular automata as a way to test them

It does not seem easy to derive, from the hypothesis of the mathematical multiverse (or actually from any other variant of the multiverse hypothesis) concrete predictions relative to our world. However, something can be done. If we start from the hypothesis that we live in a typical universe in the set of all the universes consistent with our existence −the so-called «mediocrity principle», which can be vindicated by means of statistical arguments− there are at least three assertions (proposed by Tegmark) which should take place. The first two can be formulated as follows:

---

[11] STOEGER (2007) 455.
[12] DAVIES (2007) 497.
[13] Ibídem.

«—**Prediction 1**: The mathematical structure describing our world is the most general among those consistent with our observations.

—**Prediction 2**: Our observations are the most general consistent with our existence.»[14]

The third statement is the following:

«Our future observations are the most general among those consistent with our past observations.»[15]

As in the case of the second prediction, what is being stated here is the fact that the behavior of the universe we observe cannot be more specific than what is strictly necessary to guarantee our existence. What will happen in the future should be determined only, and in the most general way, −consistent with the anthropic condition− by what has been observed in the past. Thus we should not expect that nature exhibits unnecessary regularities along time. In this way, we can consider this third prediction a particular case of the second.

The problem, anyway, is that these assertions are so general that it is not easy to see how they could be refuted by means of concrete observations about the structure of our world.

In the ideal case, a scientist doing research in this area should be able to study several universes with laws similar to ours (to a certain extent), so as to test the result of altering in some way the laws of nature. Do we really observe the most general laws consistent with our existence? Are our future observations the most general among all those consistent with our past observations? For instance, up to now everything seems to point to the fact that neither the laws nor the constants of nature in our universe change with time. Does this mean that, if they exhibited a minimal variability, they would be unable to generate structures such as us?

In the ideal case, the scientist would choose a sample of universes, some identical to ours, some with a certain variability of the laws of nature (or other variation providing those laws with a more general formulation than ours); and would find whether intelligent life becomes impossible in those universes. If this is not the case, we would be living in an universe with specially simple laws among those universe compatible with life as ours, and predictions 2 and 3 in Tegmark's proposal would be falsified.

Well, it is evident that we do not have a sample of universes to perform such a study. But perhaps we can reach the same goal in an indirect way, by studying a type of mathematical structures that can exist in multiple variations, and which generate worlds −at least in the context of the mathematical universe, where every consistent mathematical structure must be considered a world− where, depending on the rules and the initial or boundary conditions selected, complex entities may or may not appear and stay living.

Cellular automata make an interesting case of this type of mathematical structures. They consist of the following components: (1) a discrete space of dimension $n \in Z$ divided in cells; (2) a finite set of possible states for each cell; (3) a certain number (the same for all cells) of neighboring cells; and (4) a transition rule that fixes the next state of each cell as a function of its current state and the states of its neighboring cells. Time is considered a set of discrete instants, i.e. $t \in Z$.

---

[14] TEGMARK (1998) 4.
[15] TEGMARK (2007) 120.

Cellular automata are very useful for the question we are trying to research. First, because it provides us with models of «universes» regulated by rules quite easy to describe. And second −this is the main point− because, depending on the chosen transition rules −and the boundary conditions− the resulting behavior is a space where all the cells finally take the same value; or a periodic behavior emerges; or a chaotic (completely irregular) behavior; or a complex behavior: a situation containing particular configurations with special properties, which evolve and act in a way that sometimes strongly reminds the appearance and interaction of structures in our own universe.

In other words, the rules we choose determine different types of universe: monotonous, periodic, chaotic, or those capable of generating complex structures which evolve in different ways.

The most famous of these complex (sometimes called fractal) automata is the so-called «game-of-life», which we will tackle extensively in the next section. Such is the analogy between what happens in this automaton and the real world, that the «game-of-life» has been used by authors such as Daniel Dennett as an illustration of how a world ruled by a simple and strict physics may give rise to structures strongly analogous to living beings which, like living beings, should be described with a language of «intentions», «risk avoiding», «anticipation», «open opportunities», and so forth[16].

This circumstance makes of cellular automata in general, and complex cellular automata of the «game-of-life» type in particular, a set of mathematical objects key for the study of the predictions of the mathematical multiverse. The automata provide us with possible worlds which sometimes contain classes of objects analogous (in complexity) to those in our own universe, and which gives us the opportunity to investigate what happens to the complexity of those worlds when the rules of these interesting automata are made more general and complicated.

For instance, it is evident −at least to the degree of precision reached by our best instruments− that the laws and constants of nature do not experience any temporal variation in our world. Is this a necessary prerequisite for a universe to generate complex structures as ours? Or is this a case of a strange simplicity within the set of universes with rules that allow interesting structures to appear? That is, do we live, or not, in a typical universe in the set of those that generate dynamics similar to our world, as we should expect, according to the predictions of the mathematical universe?

We shall try to answer this question in the next section.

## 4. Cellular automata considered as universes. Experiments on the influence on complex structures of changes in the laws

A cellular automaton[17] (CA in short) is a set of finite deterministic automata (FDA) distributed in discrete cells along a regular grid. The inputs of the automata are the sets of states of their neighbors; the neighborhood is the same along the grid. CA can be one-dimensional (if the grid is a string of cells), bi-dimensional (when the grid is a surface), or higher-dimensional.

---

[16] See DENNETT (2003) cap.2.
[17] See NEUMANN (1966).

When the grids are finite, boundary conditions become essential. They determine, for instance, which is the left neighbor of the leftmost cell. A typical boundary condition is called periodic (or cyclic): (1-D) rows are turned into circles (their extreme cells become adjacent to each other); (2-D) rectangular grids into toroids (connecting the leftmost column to the rightmost column and the top row to the bottom row). Static (or closed) boundary conditions are also common, where the extreme cells are assumed to be connected to permanent 0-state cells.

A given CA can be tested (executed) with different initial conditions: the initial states of all the cells in the grid.

### *4.1 Experiments with one-dimensional cellular automata*

A one-dimensional CA is a linear string of cells, each containing an FDA. Each automaton in the string has the same set of n possible states; a set of neighbors, defined by the number (k) of neighboring cells to its left and to its right, which is the same for all the FDA; and a transition function (called the rule of the CA, also the same for all), which defines the next state of each automaton in the string as a function of its own current state and the states of its neighbors.

In our experiments, we have worked with one-dimensional CA with the following properties:

- Number of states: n=2, represented by 0 and 1. Usually state 0 will be called «dead» and state 1 will be called «alive».

- Maximum distance of neighbors: k=2, which means that the number of neighbors for each cell is 4 (two to the left and two to the right).

- Therefore, the next state of each cell is a function of five binary variables (the state of the cell itself and its four neighbors). The number of different possible rules is thus $2^{32}$. A given rule can be defined by a 32 bit binary string such as the following: 01001011010010110100101101001011, where each bit defines the next state of each cell for all the possible values of the five input variables in the natural order.

- An additional restriction will be imposed on the rules, to prevent spontaneous generation: when the current state of a cell is «dead» and the states of all its neighbours are «dead», the next state of the cell must be «dead». This means that the rule of the CA must always start by 0, and the number of possible rules becomes $2^{31}$.

- For our tests, we have selected a CA grid with cyclic boundary conditions.

Stephen Wolfram classified[18] one-dimensional CA into four broad categories: (i) Class 1: ordered behavior; (ii) Class 2: periodic behavior; (iii) Class 3: random or chaotic behavior; (iv) Class 4: complex behavior. The first two are totally predictable. Random CA are unpredictable. Somewhere in between, in the transition from periodic to chaotic, a complex, interesting behavior can occur.

---

[18] See WOLFRAM (2002).

Chris Langton discovered[19] that there is a relation between the λ (lambda) value of a set of rules and the Wolfram classification. He quantified the classification scheme by introducing parameter λ, and hypothesized that CA's computational capability is related to their average dynamical behavior, which λ is claimed to predict. For binary-state (1-D) CA, the Lambda parameter is the probability that a given neighborhood (chosen among all the possible configurations) leads to a «living» state, and is equal to the fraction of 1's in the rule. CA produce ordered behavior when values of lambda are close to zero or one, and chaotic behavior somewhere in between. As λ reaches a critical value (λc, the edge of chaos), rules tend to exhibit long-lived, complex behavior.

Since the total number of rules is too large to allow for a systematic study, we have selected at random four automata with rules which generate a complex behavior, and explored what happens when one or two mutations are applied to these rules. Tables 1 and 2 show the results. The Complex/Chaotic corresponds to the case when the modified CA displays a complex behavior for some initial conditions, and a chaotic behavior for other.

**Table 1. What happens when all possible single mutations are applied to a CA with complex behavior.**

| Rule | Complex | Chaotic | Complex /Chaotic | Ordered/ Periodic | Total |
|---|---|---|---|---|---|
| 01001011010010110100101101001011 | 21 | 1 | 0 | 9 | 31 |
| 01010110011011101110111010000000 | 9 | 20 | 2 | 0 | 31 |
| 01100110011001100110011001100110 | 9 | 14 | 3 | 5 | 31 |
| 00111100001111000011110000111100 | 10 | 12 | 2 | 7 | 31 |

**Table 2. What happens when all possible consecutive double mutations are applied to a CA with complex behavior.**

| Rule | Complex | Chaotic | Complex /Chaotic | Ordered /Periodic | Total |
|---|---|---|---|---|---|
| 01001011010010110100101101001011 | 14 | 3 | 2 | 11 | 30 |
| 01010110011011101110111010000000 | 3 | 26 | 1 | 0 | 30 |
| 01100110011001100110011001100110 | 0 | 17 | 8 | 5 | 30 |
| 00111100001111000011110000111100 | 5 | 12 | 1 | 12 | 30 |

---

[19] See LANGTON (1990).

These tables show that the behavior of the automata changes from complex to ordered, periodic, or, most frequently, chaotic, when its rule (i.e. one of the laws of nature for a world regulated by its rule) is modified; but in a significant number of cases, the automaton maintains a complex behavior after the change.

The next tests were performed on the second CA in tables 1 and 2. This CA has been designed with a rule similar to the laws of our universe, according to the following considerations:

1. The rule is symmetric, i.e. the neighbors to the left have the same effect as the neighbors to the right.
2. Too few neighbors or too many neighbors tend to cause the «death» of the central cell (as in the Game of Life).
3. About one half situations cause cells to become «alive», the other half make them «dead». According to Langton, this makes complex behavior more probable.

As indicated in the tables, the rule for this automaton is the following: 010101100110111011101110000000. Figures 1, 2 and 3 show the evolution of this CA for three different initial conditions. In these figures and the following ones, time is the vertical axis, «dead» cells are shown in white, and «living» cells in black.

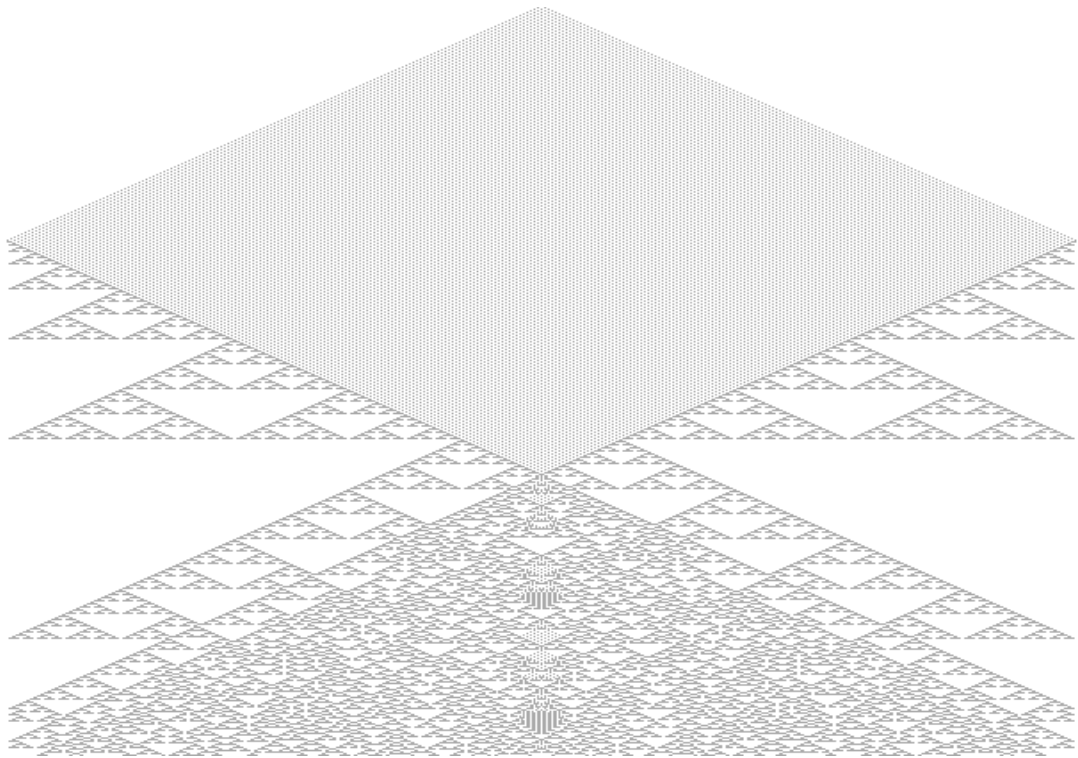

**Figure 1: Complex behaviour of the CA with symmetric rule 010101100110111011101110000000 with initial conditions (300x0),1,(300x0).**

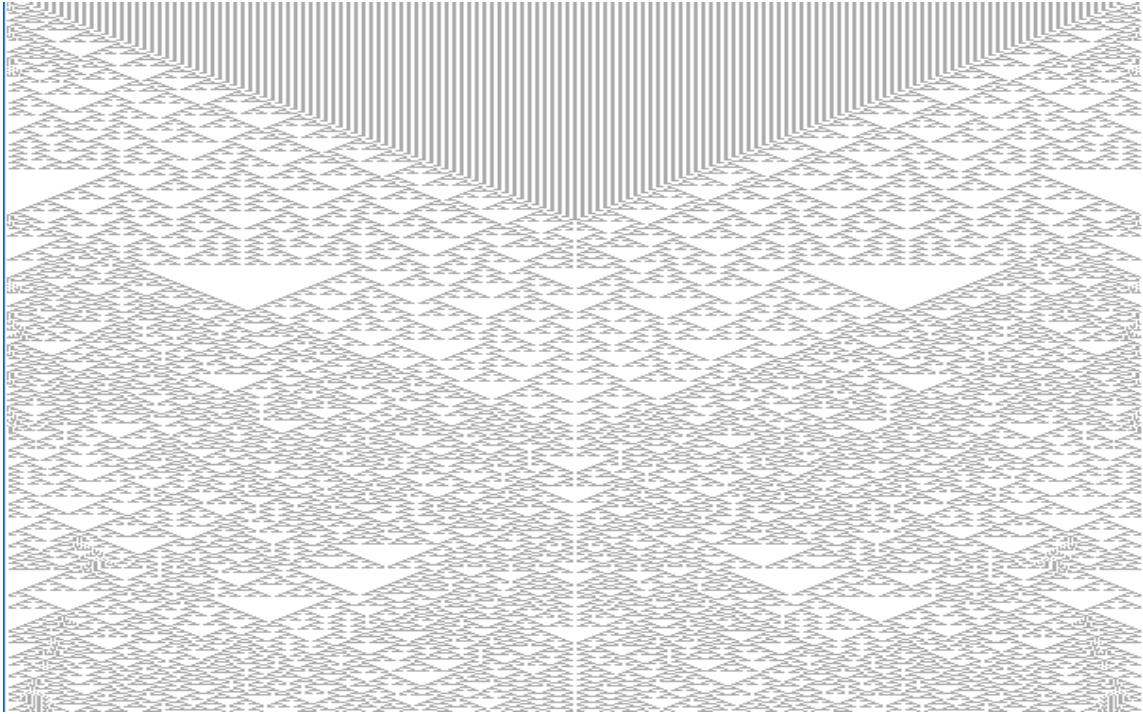

**Figure 2: Complex behaviour of the CA in Figure 1 with initial conditions 1100 repeated to make a 601 long string.**

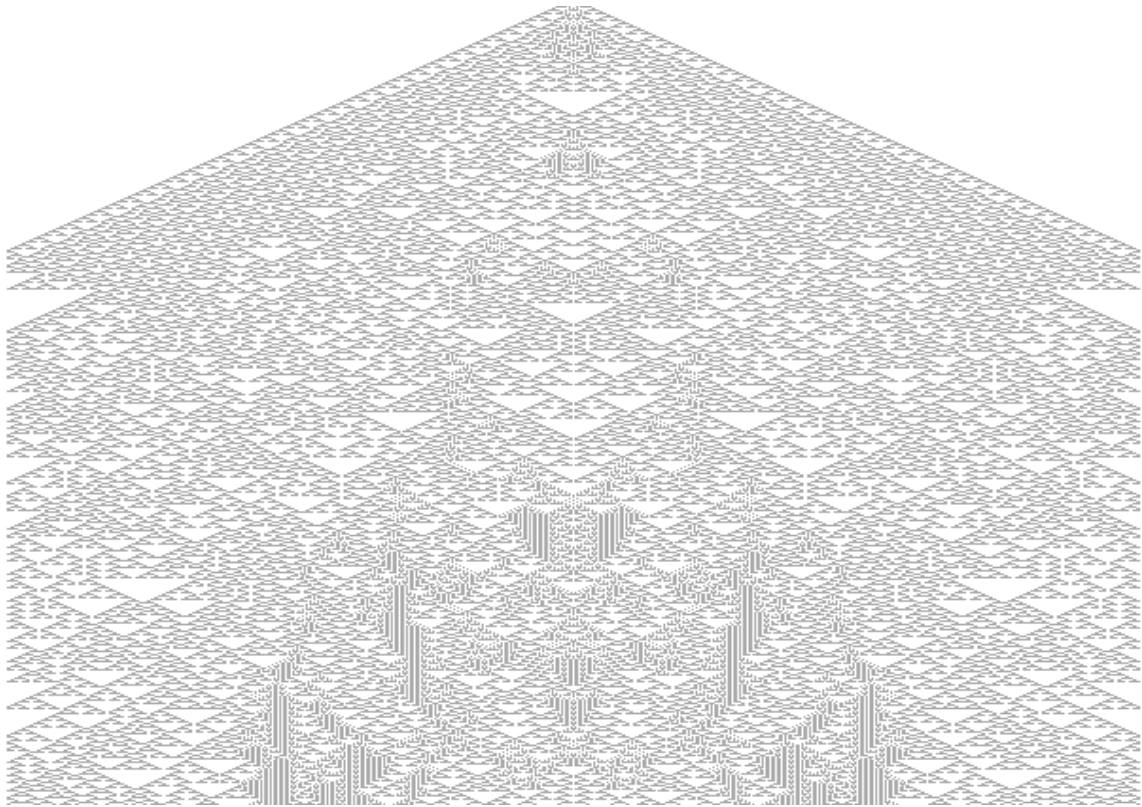

**Figure 3: Complex behaviour of the CA in figure 1 with initial conditions (299x0)111(299x0)**

If the rule of this CA is applied a single mutation, symmetry is lost, and the complex behavior for these initial conditions becomes chaotic (see fig. 4-6).

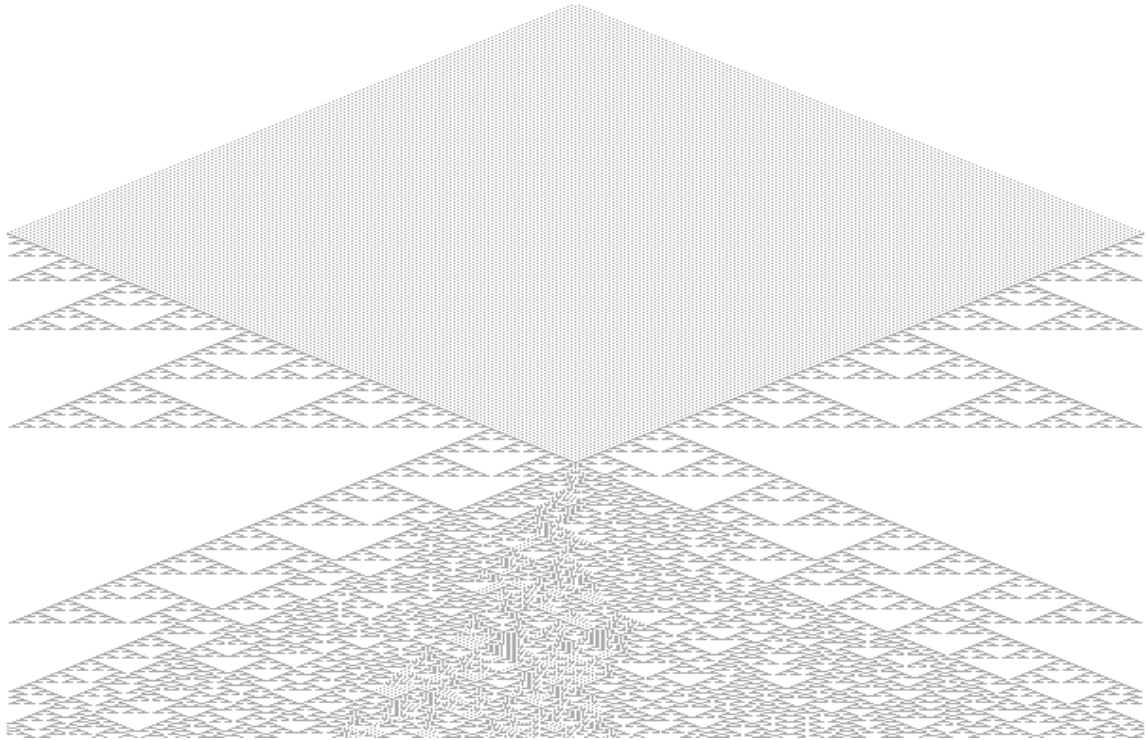

**Figure 4: Chaotic behaviour of the CA with asymmetric rule 010101100010111011101110100000000 with initial conditions (300x0),1,(300x0).**

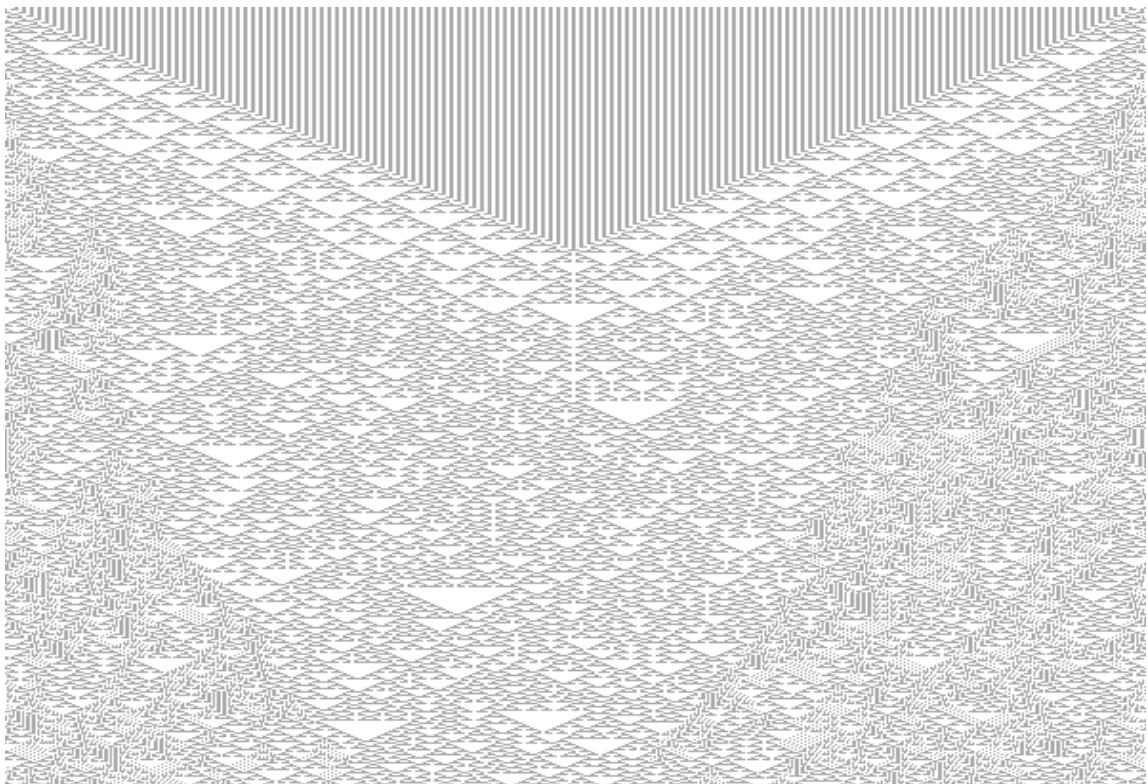

**Figure 5: Chaotic behaviour of the CA in Figure 4 with initial conditions 1100 repeated to make a 601 long string.**

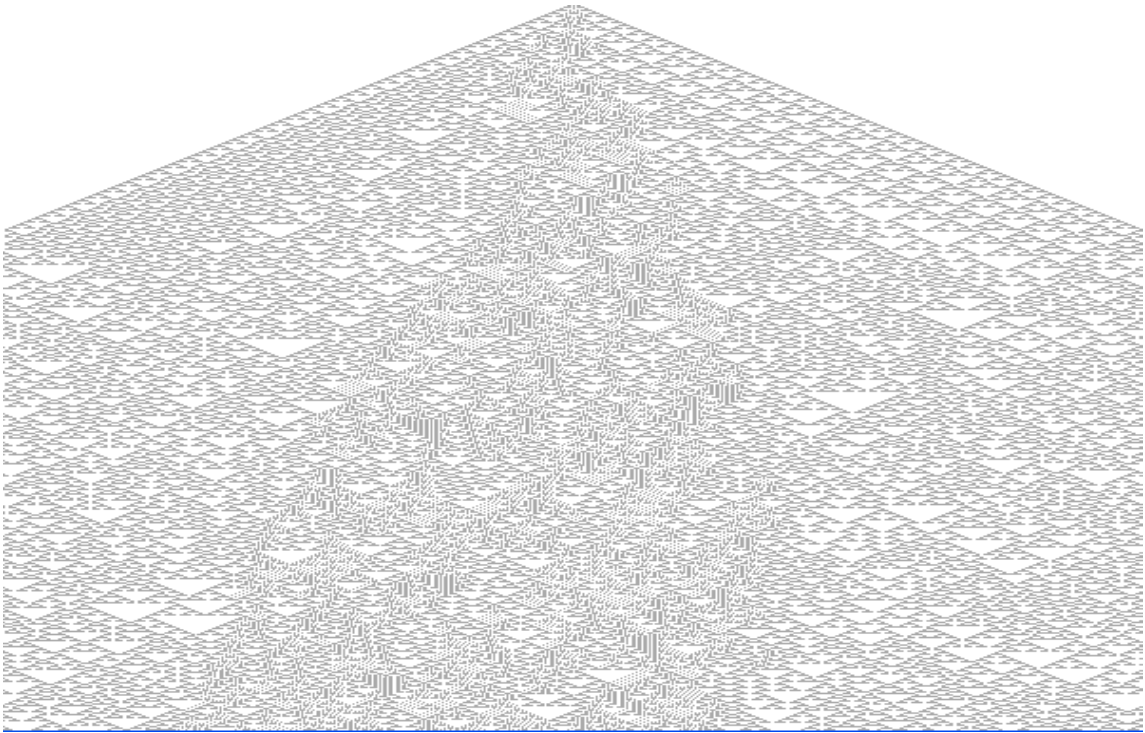

**Figure 6: Chaotic behaviour of the CA in Figure 4 with initial conditions (299x0),111,(299x0).**

However, if we apply a double mutation which maintains symmetry, the behavior of the CA will still be complex (see figures 7-9).

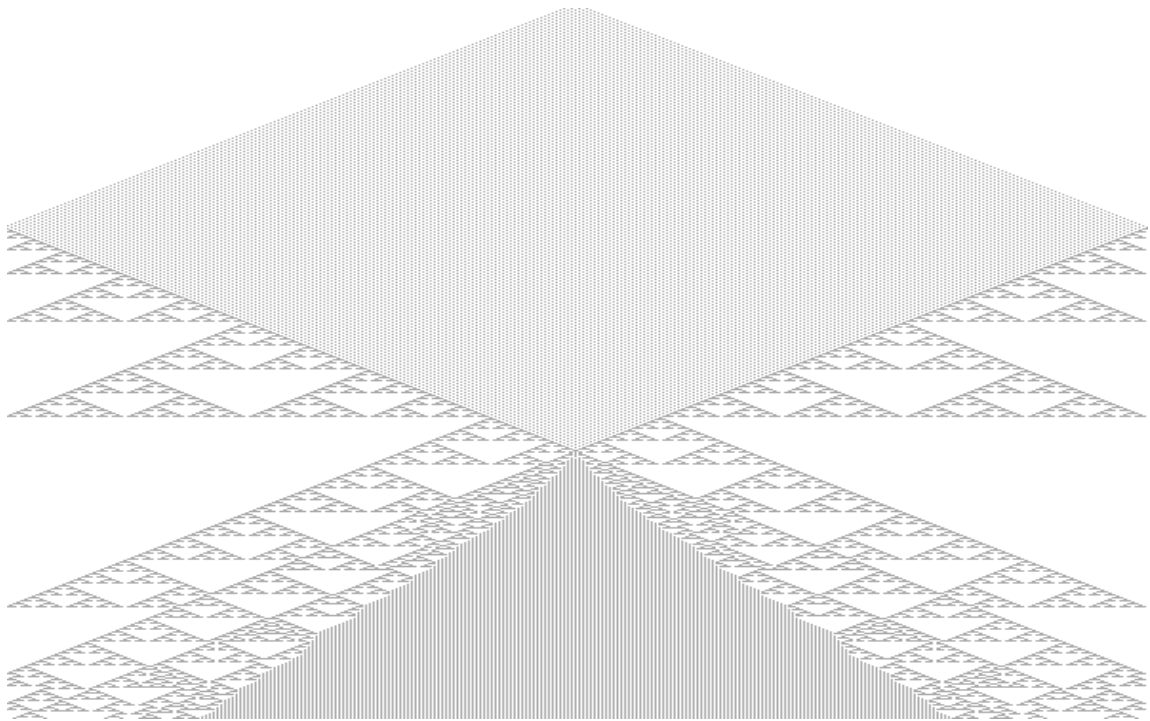

**Figure 7: Complex behaviour of the CA with symmetric rule 010101100000111011101110100000000 with initial conditions (300x0),1,(300x0).**

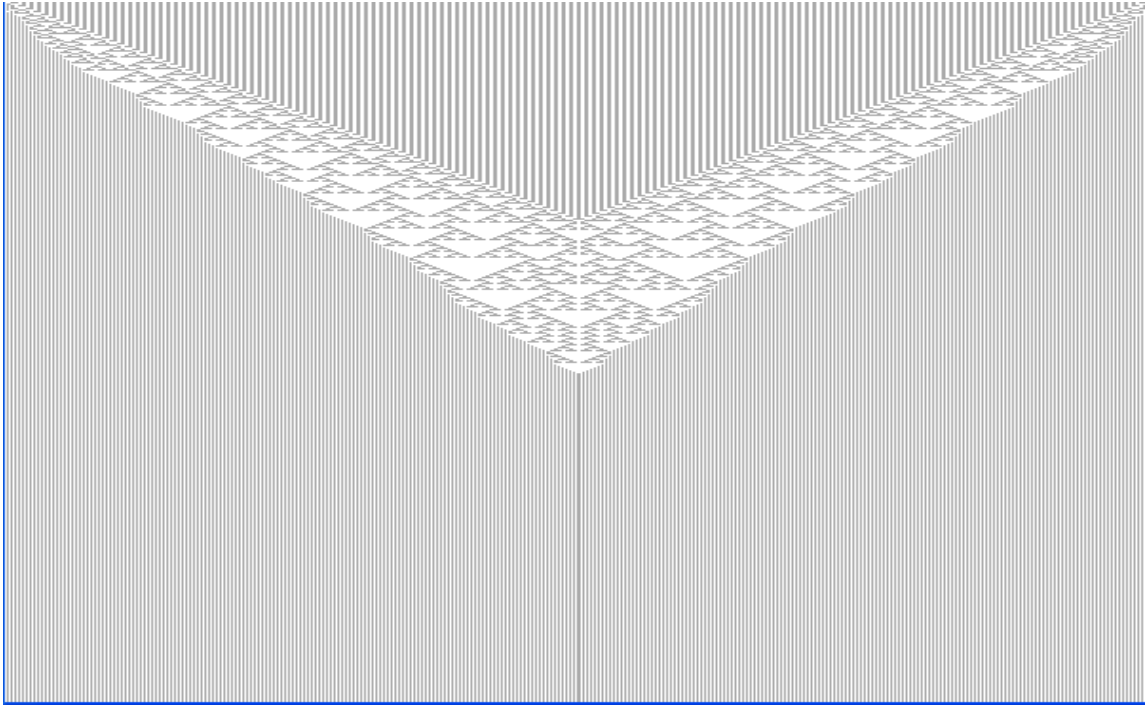

**Figure 8: Complex behaviour of the CA in Figure 7 with initial conditions 1100 repeated to make a 601 long string.**

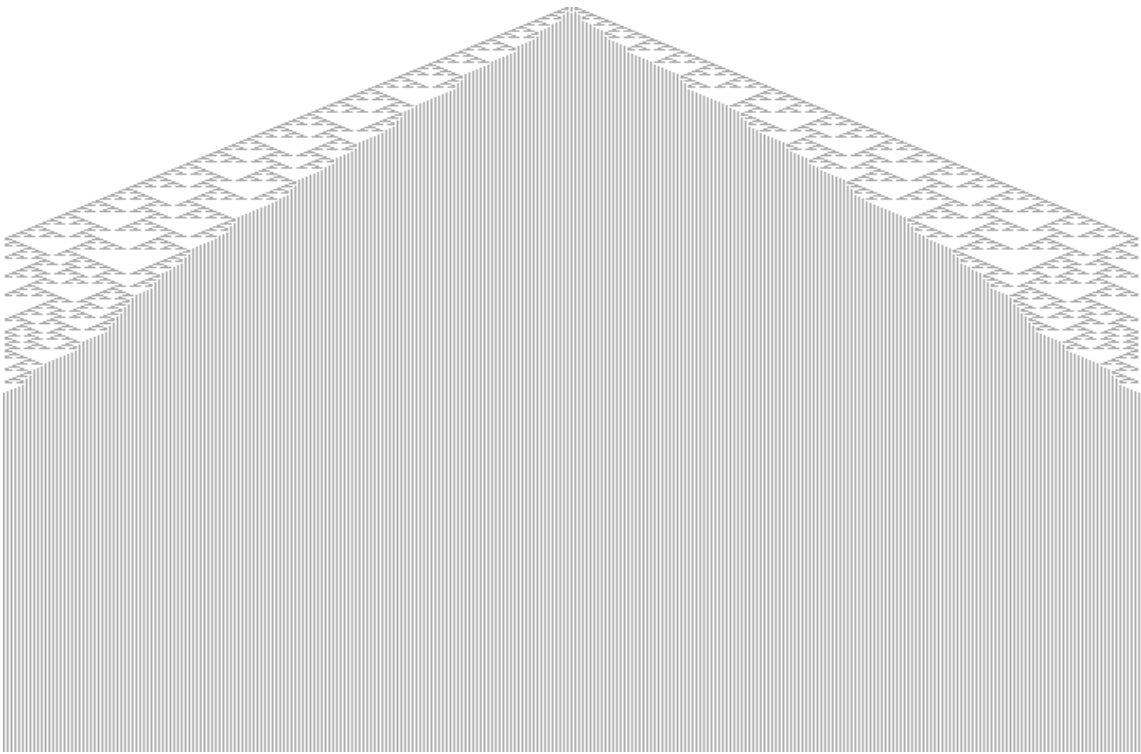

**Figure 9: Complex behaviour of the CA in Figure 7 with initial conditions (299x0),111,(299x0).**

The same happens with other double mutations which also keep the symmetry.

We next complicated even more the «laws of nature» in this universe, applying the original rule for some time during the evolution of the CA, then applying a double mutation for some time, and letting the CA go back to its original rule. In general, we got a complex behavior, even though the evolution in each case is visibly different from the corresponding histories of the original CA (those shown in figures 1-3). Figures 10-13 show the results.

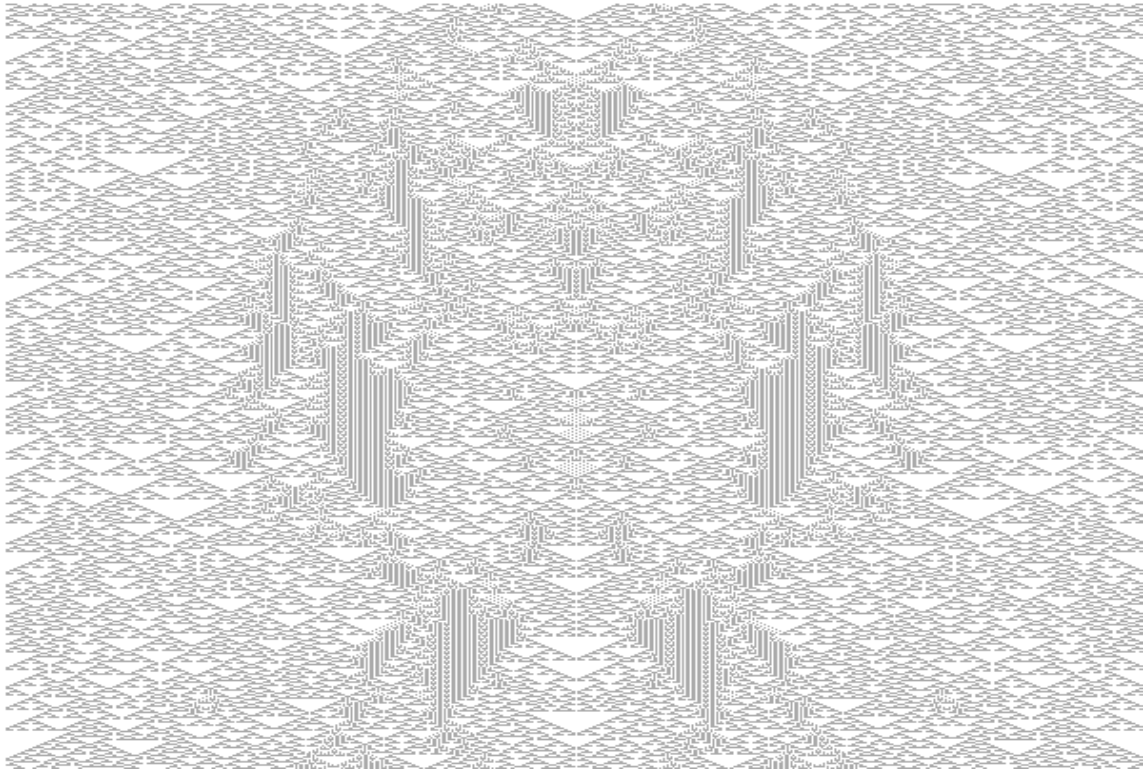

**Figure 10: Complex behaviour of the CA with symmetric rule 010101100110111011101110100000000 with initial conditions (300x0),1,(300x0) during generations 250-750.**

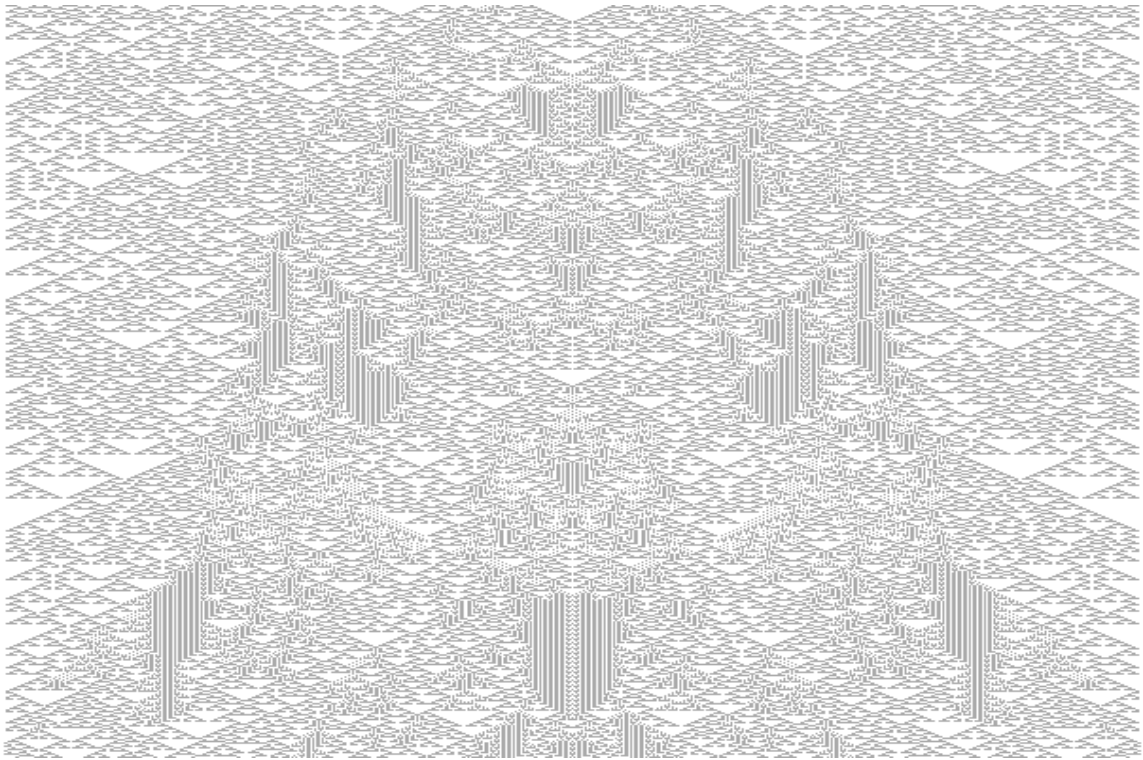

**Figure 11: Complex behaviour of the CA in figure 10 with double mutation RULES[10 19]←0 applied in generation 500 and undone in generation 600.**

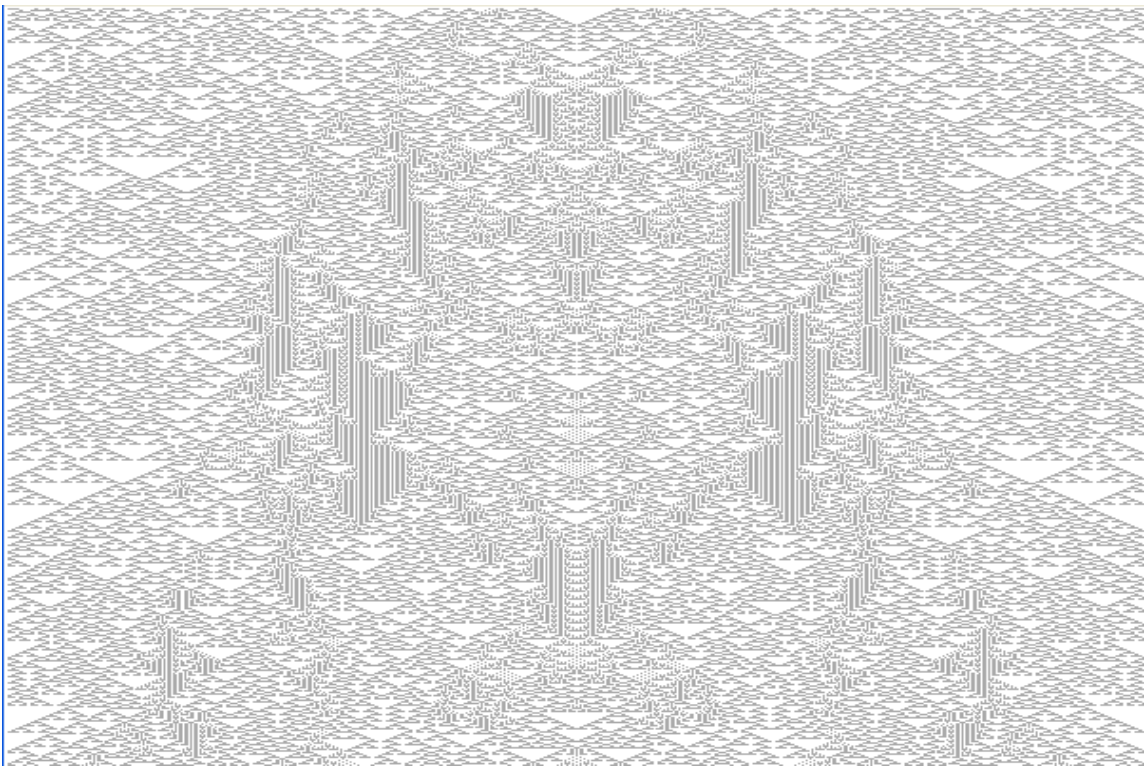

**Figure 12: Complex behaviour of the CA in figure 10 with double mutation RULES[10 19]←0 applied in generation 500 and undone in generation 510.**

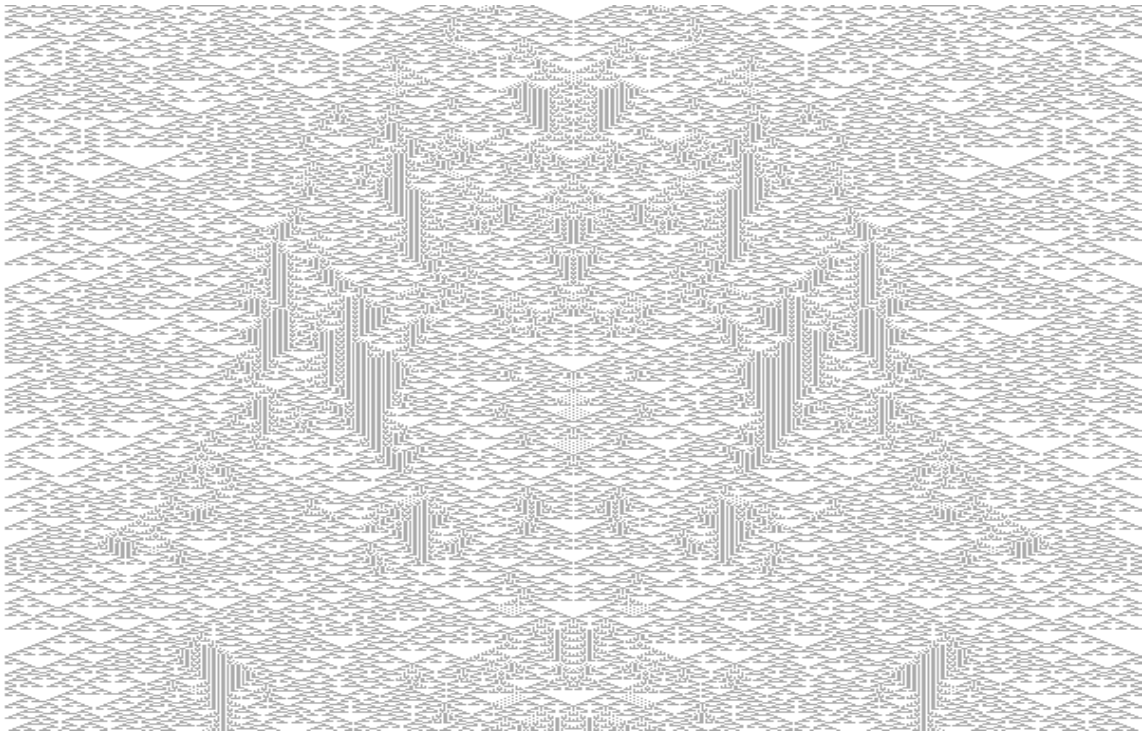

**Figure 13: Complex behaviour of the CA in figure 10 with double mutation RULES[10 19]←0 applied in generation 500 and undone in generation 501.**

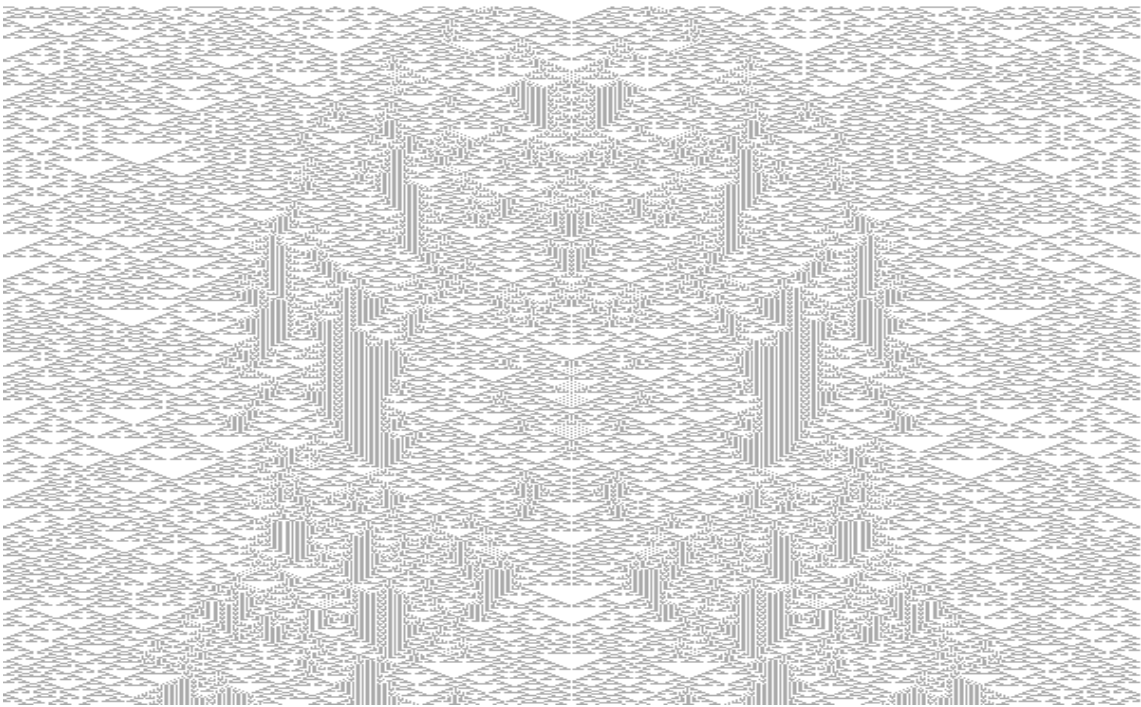

**Figure 14. A run of the original cellular automaton with double mutation applied randomly at every step.**

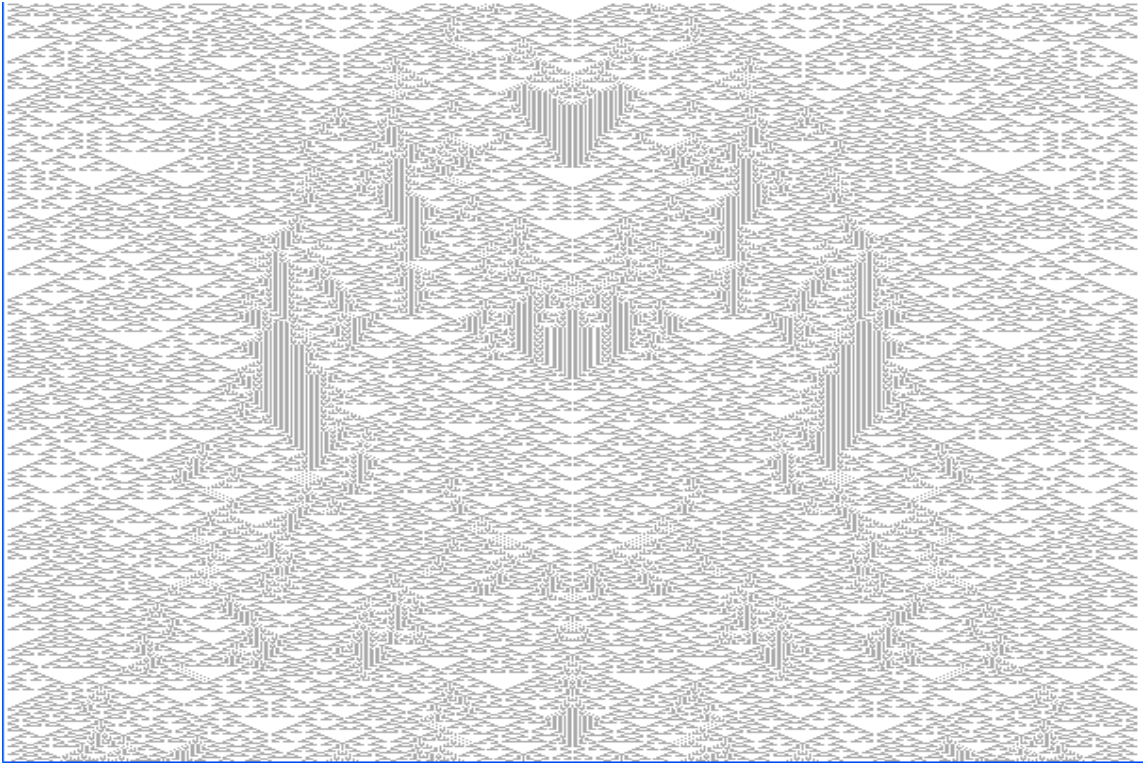

**Figure 15. Another run of the original cellular automaton with double mutation applied randomly at every step**

Finally, the double mutation RULES[10 19]←0 was applied in every generation with a probability of 1 in 1000. Figures 14-15 show two different runs of this process. These experiments show that the CA-universes which exhibit an interesting behavior (i.e. a complex behavior) do not lose it automatically if they suffer certain changes in their rules. This suggests that the most general form of the laws of nature allows for a certain time variability in those laws.

In any case, as shown by the above figures, the actual histories of the modified universes will be different. Also, from so a general perspective, we cannot say what will happen to a given structure in a complex world when its laws are complicated with changes of any type (for instance, by introducing temporal changes in the laws, as shown in the experiments).

But we can certainly expect at least that many variations of such universes will allow the existence of entities of analogous complexity.

### *4.2 Experiments with bi-dimensional cellular automata*

To study what happens to particular structures in a complex world when changes in the laws occur, we decided to analyze bi-dimensional automata of the Game-of-Life type[20], one of the best explored up-to-now.

---

[20] See WOLFRAM (1986).
21

So far, CA have proved very powerful to simulate many real life applications and phenomena. It has also been proved that some 1-D and 2-D CA, such as the Game-of-Life (Life, in short), are computationally equivalent to the Universal Turing Machine[21].

The 2-D CA called the Game-of-Life was designed by John Conway. It consists of a matrix of cells, where each cell may take one of two states: alive and dead (respectively represented by one and zero). Each cell has eight neighbors, according to the Moore neighborhood (in the eight main directions of the compass). At every time step, also called a generation, each cell computes its new state by determining the states of the cells in its neighborhood and applying the transition rules to compute its new state. Every cell uses the same update rules, and all the cells are updated simultaneously. The next state of a cell is determined by the rule B3/S23, which means that cells are born (go from the dead to the living state) if they have exactly three living neighbors, and survive if they have two or three living neighbors. In all other cases, a cell dies or remains dead.

Different variants of the game of life have been defined. HighLife, for instance, differs because its rule is B36/S23 (i.e. a cell is also born if it has 6 living neighbors). Life-3-4 has the rule B34/S34 (cells are born or survive if they have 3 or 4 living neighbors). Seeds is a CA with the rule B2/S (a cell is born if it has exactly two living neighbors, but it never survives).

Formally, a cellular automaton (CA) «universe» of this type consists of two things:

- The CA definition, consisting of:

    o A set of states, in our case restricted to {0,1}

    o A space dimension, in our case 2

    o A space size, in our case 60x60

    o The space boundary condition, in our case a flat toroid

    o A neighborhood, in our case the Moore neighborhood

    o The CA rule, which describes the way in which the initial conditions will change with time. The rule may be selected from a large set of possible rules: Life, HighLife, 3-4-Life or Seeds, among others. Mixed rules will also be used in some cases. For instance, the Life rule up to generation 25; the HighLife rule up to generation 50; and so on, periodically. Other mixed cases will also be considered.

- The initial condition: a matrix of initial states (in our case 60x60) with values in the set of states (in our case {0,1}).

In our experiments, we first select a given CA definition (i.e. a rule, since all the other parameters of the CA are fixed). We have developed a genetic evolution program which selects some of the most «interesting» initial conditions for that CA definition: those which give rise to a good number of interesting small CA structures, specially gliders (which make it possible to design logical gates, and thus provide Life with the capability of universal computation), but also r-pentominos, or exploders.

---

[21] See SARKAR (2000).

The genetic evolution program has the following additional parameters[22]:

- Population size: 64 random initial conditions for the CA whose rules have been chosen.

- Size of the original population random initial conditions: 30x30 centered on the 60x60 universe. The remainder of the CA space is set at state 0 (dead cells).

- Number of evolutionary steps: 400

- The fitness score for each member of the population is the number of appearances of «interesting» structures during generations 40 to 54 during the execution of the CA.

- Population replacement after each step: the 16 individuals with least scores are replaced by another 16, obtained from the 16 with the highest scores (which are paired randomly) by means of sexual reproduction, which uses the genetic operations of genetic recombination and random mutations to generate new genomes.

- After the indicated number of generations, the program returns that universe which obtained the maximum score in the whole process (the best initial conditions for the chosen CA rule). This CA is analyzed using the Golly application, unloaded from http://golly.sourceforge.net/ to find all the interesting structures it generates.

Table 3 shows the results obtained in our experiments, seven for each type of rule (Life, HighLife, and a periodic mixed Life/HighLife rule, as described above), selecting for both gliders and r-pentominos. For comparison, row 4 in this table shows what happened when the evolutionary algorithm was changed to select for a different type of object (two simple types of exploders). Five experiments of the latter type were performed.

**Table 3. Summary of experiments as a function of rule type**

| Type of rules | Average life length | Gliders | R-Pent. | Exploders |
|---|---|---|---|---|
| Life | 578 | 106(32.78+4x∞) | 22 | many |
| HighLife | 386 | 33(57.9) | 2 | 24 |
| L→HL→L→... | 552 | 59(57+4x∞) | 13 | 40 |
| Life (exploders selected) | 938 | 79(24.04+∞) | 26 | 88 |

---

[22] See more technical details of this program and the related experiments in our paper: ALFONSECA, M. and Soler Gil, F. J. (forthcoming), «Evolving interesting initial conditions for cellular automata of the Game-of-Life type».

A few conclusions derived from Table 3:

- Interesting behavior appears both with Life, Highlife and the mixed rule. The highest score corresponds to one experiment which used Life, and the highest number of gliders appeared in another experiment also using Life, but the second and fourth highest on both accounts (respectively) was reached by one experiment which used HighLife. Sometimes, however, for all rules, the evolution experiment does not generate much interesting behavior. This happened in 6 out of 7 HighLife experiments, 4 out of 7 mixed experiments, and 3 out of 7 Life experiments, which seems to indicate that the rules of Life are somewhat more prone to the appearance of «interesting» behavior than the rules of Highlife, while the mixed rules occupy an intermediate position.

- The life length of an experiment is considered to end when the CA configuration goes into a static situation, where the states of all the cells remain the same forever (not necessarily dead), or a periodic configuration, where the states of the cells oscillate with a certain period.

- Gliders are generated much more frequently than r-pentominos. Their column in Table 3 shows two numbers: the first is the total number of different gliders generated by the CA with the evolved initial conditions for all the experiments associated to a given rule; the second provides the average life of the gliders (the average number of generations that they endured, computed as the sum of the number of generations that each glider exists divided by the number of gliders, excluding permanent gliders, if any). In some experiments, two gliders collided and destroyed one another. In many cases, a glider is generated and destroyed almost immediately. In some cases, one or more gliders survived permanently, giving rise to a final periodic configuration with a period of 240 generations[23].

- When exploders were not selected for, they appear anyway, relatively frequently. Sometimes they are quite complex (see figure 16, which also displays a glider). Similarly, when exploders are selected for, rather than gliders and r-pentominos, the latter appear anyway. It is interesting to notice that the average number of gliders per experiment (15.8) and their duration (24.04) are similar to those obtained when gliders are directly selected using the game of Life rules (15.14 / 32.78), while three times less gliders (4.7), but with longer duration (57.9), appeared when gliders were selected with the HighLife rules. On the other hand, the experiments where exploders are selected had a higher average life length (about double that of the previous experiments).

---

[23] It takes a glider 4 generations to move a position diagonally. So, in an universe of size 60x60, the period of a permanent glider will always be 240.

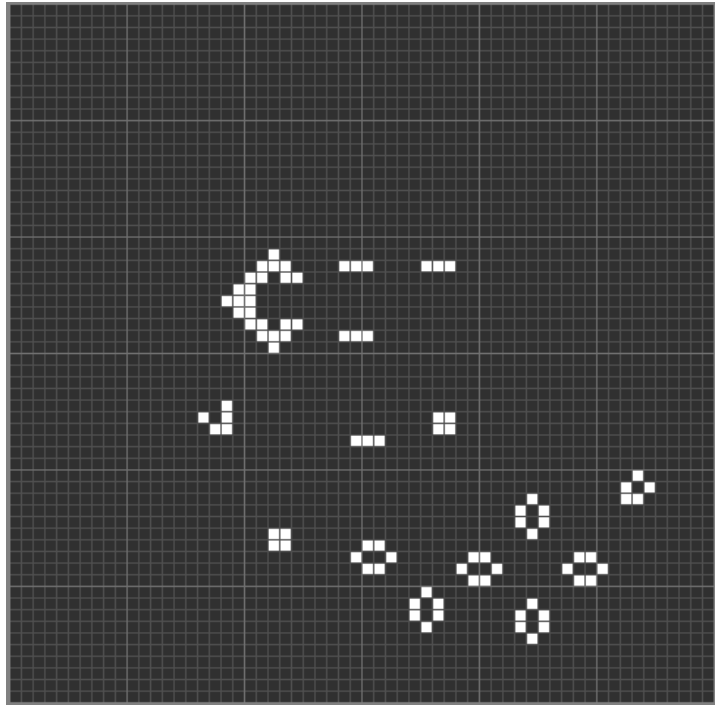

**Figure 16. An exploder at generation 239 in experiment AA8B. A glider is also visible.**

In the next set of experiments, we have tried to find out the effect of changing the rules, but maintaining the initial conditions developed in the previous set of experiments. In this way, we used the initial conditions evolved for Life with the HighLife rules and vice versa. The results are shown at rows 3 and 4 in table 4. For comparison, rows 1, 2 and 9 show the average results for the CA summarized in Table 3. It can be seen that, when the rules for HighLife are used with the initial conditions for Life, the results are «less interesting» (shorter life length, and very few gliders and similar objects appear). In the opposite case, however, the situation is not so clear: we get a longer life length, but a similar (even slightly larger) number of objects. Again, as in previous experiments, the HighLife rules are seen to be somewhat less versatile than the Life rules.

**Table 4. Summary of experiments when initial conditions evolved for Life are used with the HighLfe rules and vice versa**

| Rules evolved for A, used with B | Average life length | Gliders per exper. | R-Pent. per exper. | Exploders per exper. |
|---|---|---|---|---|
| Life / Life | 578 | 15.1 | 3.1 | many |
| HighLife / HighLife | 386 | 4.7 | 0.3 | 3.4 |
| Life / HighLife | 170 | 1.1 | 0.6 | 2.4 |
| HighLife / Life | 961 | 20.3 | 5.3 | 8.9 |
| Life / Life→HighLife | 394 | 5.3 | 1.6 | 2.3 |
| HighLife / HighLife→Life | 1827 | >16.7 | 2.9 | many |
| Life / L→HL→L | 791 | 14.4 | 3.9 | 5.9 |
| Life / L→HL→L→... | 478 | 9 | 2 | 4.1 |
| L→HL→L→... / L→HL→L→... | 552 | 8.4 | 1.9 | 5.7 |

The next set of experiments tried to find the effect of changing the rules during the execution of one of the automata generated in the previous examples. In this case, the initial conditions evolved for one type of laws (Life or HighLife) were applied to a CA which runs under those laws until generation 46, then changes to the opposite laws during the remainder of its «life». Thus, if the automaton was generated using the rules of Life, at generation 46 the rules would be changed to HighLife, and vice versa. The results can be seen at rows 5 and 6 in Table 4. From their observation we may get the following conclusions:

- The mixed rule of the form Life→HighLife (with initial conditions evolved for Life) generated a less complex behavior (shortest life, less gliders and other interesting objects) than the equivalent experiments where the rules of Life were allowed to apply always, but a slightly more complex behavior than those where the rules of HighLife applied always, with initial conditions evolved for HighLife, and significantly more complex than those where the rules of HighLife were applied with initial conditions evolved for Life.

- The mixed rule of the form HighLife→Life (with initial conditions evolved for HighLife) generated a behavior at least as complex (in fact slightly more complex) than those where initial conditions evolved for Life were applied to a CA running with the Life rules.

In the next set of experiments of this type, we started with CA with rules of the Life type and let it develop for 46 generations; then changed the rules to HighLife, executed them for 4 generations, and restored again the rules to Life for the remainder of their development. The initial conditions evolved for CA with Life rules where applied to these CA. Row 7 in table 4 show the results: a CA of this type performs comparably as that with the Life rule.

In the next set of experiments, we started with CA with rules of the Life type and let it develop for 46 generations; then changed the rules to HighLife, executed them for 4 generations, and restored again the rules to Life. This procedure was repeated periodically every 50 generations. The initial conditions evolved for CA with Life rules where applied to these CA. Row 8 in table 4 show the results: a CA of this type performs slightly worse than that with the Life rule, but about the same as the CA with periodic rules and initial conditions evolved for them (row 9 in the table).

To end this analysis, we decided to perform a few experiments using completely different types of cellular automata:

- The first we tried was Life-3-4, defined by the rule B34/S34. It resulted not to be amenable to this kind of experiments: there are no small long-lived structures similar to gliders, and therefore evolutionary algorithms do not seem to work; they fail to improve the best score, which is typically reached randomly in the first generation at a very low value, and remains there for the remainder of the evolutionary process.

- Then we tried Seeds, a CA defined by the rule B2/S. With this apparently radical rule, however, it is possible to generate structures similar to gliders which move one step in a certain direction from one generation to the next. We selected for this structure in our evolutionary process. This CA has the problem that the number of cells alive increases quickly (in our experiments this happened always before the 100th «generation»), finally covering about 20% of the available space, and their distribution is more or less chaotic, which produces the effect that any glider that may appear from this point on will be quickly smothered by the neighboring cells and will stay there for just one or two generations. This chaotic behavior seems to stay forever, which means that the CA never reaches a static or periodic situation. To reduce this effect, we restricted the «life» of the CA to the first 60 generations and counted the number of gliders which appeared, and their duration. In the two experiments performed, 6 and 7 gliders were produced (respectively) with an average duration of 11.4 generations.

- Finally we tried the following mixed case: we started with CA of the Life or the HighLife type, and let them develop for 46 generations; then we changed the rules to 3-4-Life, executed them for 4 generations, and restored the rules to Life or HighLife. Table 5 shows the results. In some cases, the automata could not recover their former complexity after the change: not a single glider was produced after the original rule was restored. In other cases, however, new gliders were generated. We can conclude that even this drastic change in the rules decreases only moderately the average complexity of the development, i.e. sometimes the complexity of a given experiment is destroyed, but in other cases the CA is able to recover and proceeds to generate new complex behavior for a reasonable number of generations.

Table 5. Experiments with Life/3-4-Life mixed rules

| Rules evolved for A, used with B | Life length | Gliders | R-Pent. | Exploders |
|---|---|---|---|---|
| Life / Life→Life-3-4→Life | 341 | 9.6 | 3.9 | 3.4 |
| HighLife / HighLife→Life-3-4→HighLife | 260 | 5.6 | 2.3 | 4 |
| Life / Life | 578 | 15.1 | 3.1 | many |
| HighLife / HighLife | 386 | 4.7 | 0.3 | 3.4 |

Our conclusion: CA with mixed (time-dependent) rules are fully as capable to generate interesting behavior as those with «pure» rules, in some cases even more so.

### 4.3 Additional experiments with bi-dimensional cellular automata

It has been mentioned before (in the preceding sub-section) that a combination of gliders and glider generators and destroyers makes it possible to design logical gates, and thus provide Life with the capability of universal computation. Since any possible computation can be built (in principle) as a combination of just three types of possible gates (OR, AND and NOT), with a large enough grid, any possible computation can be performed by the game of Life.

We have tested the structures that perform the three logical gates in Life, and have found out that all of them are compatible and stable with four different cellular automata rules:

- Life: B3/S23
- B38/S23
- B3/S238
- B38/S238

Therefore, these four types of cellular automata are computationally complete. What is more, any temporal combination of those four automata will also be computationally complete and all computational structures which can be built will be stable and functional in them. We have confirmed this by testing just two periodic situations where the rules for Life and B38/S23 on the one hand, and Life and B38/S238 on the other, alternate every 50 generations. Therefore we have found an infinite set of cellular automata rules (4 basic ones, with a constant rule, and an infinity of time-dependent combinations) all of which are capable of universal computation (in our parlance, compatible with life). In this case, the probability of finding one cellular automaton in this family with constant rules is mathematically equal to zero (4 divided by infinity).

## 5. Discussion of the results

In this paper, we have checked on the fact that the only version of the multiverse that could be a suitable candidate for explaining the fine tuning of the laws of our universe to make the existence of complex entities in general and of intelligent beings in particular possible is Tegmark's mathematical multiverse. Then we have focused on the question whether the peculiar laws of our universe can be explained from the hypothesis that ours is a typical «complex-universe» or «life-enabling universe» among the set of all the worlds which includes all the consistent mathematical structures.

In order for this to be the case, the laws of our universe should be the most general among those consistent with our existence. To test this, in the previous section we have analyzed and proved that the most general form of those complex universes (which must exist, according to Tegmark's hypothesis) whose structure corresponds to that of cellular automata, are those whose rules exhibit some kind of temporal variability.

If this result can be extrapolated −and we should not forget that there are numerous authors from Martin Gardner to Daniel Dennett who have suggested the existence of a very close relation between the «game of life» and our universe−, it would imply that our universe is not typical at all, since it attains a high degree of complexity with laws and physical constants specially simple, as they are not a function of time. On the other hand, it is obvious that the number of possible universes compatible with life which would exhibit some kind of temporal dependence in their laws and physical constants, while keeping within the allowable margins of values, must be infinitely more probable than those of our type, which means that the probability of our existence in a world such as ours would be mathematically equal to zero.

In consequence, the results presented in this paper can be considered as an inkling that the hypothesis of the multiverse, whatever its type, does not offer an adequate explanation to the peculiarities of the physical laws in our world. Multiverses are either too small or too large to explain fine-tuning. All the multiverses which have been proposed, except the «mathematical multiverse», are too small, so that they merely shift the problem of fine tuning up one level. But the «mathematical multiverse» is too large, in the sense that, in its context, the simplicity of our world becomes inexplicable.

## References

BARROW, J. - TIPLER, F. (1986), *The Anthropic Cosmological Principle* (Oxford, Oxford University Press).

COLLINS, R. (2003), «The evidence for fine-tuning». In: N. Manson (ed.), *God and Design: The Teleological Argument and Modern Science* (London, Routledge).

DAVIES, P. (2007), «Universes galore: where will it all end?», In: B. CARR (ed.), *Universe or multiverse?* (Cambridge, Cambridge University Press).

DENNETT, D. (2003), *Freedom evolves* (New York, Viking Penguin).

ELLIS, G. F. R. – KIRCHNER, U. – STOEGER, W. (2003), «Multiverses and physical cosmology.» Available at: arXiv (astro-ph) 0305292.

LANGTON, C. (1990), «Computation at the edge of chaos: Phase transitions and emergent computation» *Physica D*, Volume 42 (1990) 12–37, Issue 1-3.

NEUMANN, J. VON (1966). *Theory of self-reproducing automata*, edited and completed by A. W. Burks (Urbana, IL, University of Illinois Press).

OBERHUMMER, H. - CSÓTÓ, A. - SCHLATTL, H. (2000): «Fine-tuning of carbon based life in the universe by triple-alpha process in red giants». *Science* 289(5,476) (7.7.2000) 88-90.

SARKAR, P. (2000), «A brief history of cellular automata» *ACM Computing Surveys (CSUR)* Vol. 32, No.1, (2000) 80 – 107.

STOEGER, W. (2007), «Are anthropic arguments, involving multiverses and beyond, legitimate?», In: B. CARR (ed.), *Universe or multiverse?* (Cambridge, Cambridge University Press).

STOEGER, W. – ELLIS, G. F. R. – KIRCHNER, U. (2004), «Multiverses and cosmology: Philosophical issues». Available at: arXiv (astro-ph) 0407329

TEGMARK, M. (1998), «Is "the Theory of Everything" Merely the Ultimate Ensemble Theory?» *Annals of Physics* 270 (1) (1998) 1–51.

TEGMARK, M. (2004), «Parallel universes». In: J. BARROW – P. DAVIES – C. HARPER (eds.), *Science and ultimate reality* (Cambridge, Cambridge University Press) 483.

TEGMARK, M. (2007), «The multiverse hierarchy». In B. CARR (ed.), *Universe or multiverse?* (Cambridge, Cambridge University Press).

TORBEY, S. (2007), *Towards a Framework for Intuitive Programming of Cellular Automata*, (Unpublished Master Dissertation, Queen's University, Kingston, Ontario, Canada).

WOLFRAM, S. (1986). *Theory and applications of cellular automata*, (1st ed.), (Singapore, World Scientific).

WOLFRAM, S. (2002), *A New Kind of Science* (Wolfram Media).